\documentclass[showpacs,superscriptaddress,amsmath,amssymb,nofootinbib,twocolumn]{revtex4-2}
\usepackage{graphicx}
\usepackage{epsfig}
\usepackage{overpic}
\usepackage{dcolumn}
\usepackage{bm}
\usepackage{lineno}
\usepackage{xspace}
\usepackage{multirow}
\usepackage{epstopdf}
\usepackage{xcolor}
\usepackage{soul}
\soulregister{\cite}7
\soulregister{\ref}7
\soulregister{\st}7
\usepackage[colorlinks,allcolors=blue]{hyperref}
\usepackage{verbatim}
\usepackage{enumitem}
\usepackage{todonotes}
\usepackage{subcaption}
\usepackage{dcolumn}% Align table columns on decimal point
\usepackage{bm}% bold math
\usepackage{threeparttable}
\usepackage{multirow}
\usepackage[figuresright]{rotating}
\usepackage{makecell}
\usepackage{verbatim}
\usepackage{slashed}
\usepackage{mathtools}
\usepackage{tikz-feynman}
\usepackage{utfsym}
\usepackage{amsmath}
\include{def-com}
\newcommand{\scnt}{\affiliation{Southern Center for Nuclear-Science Theory (SCNT), Institute of Modern Physics, Chinese Academy of Sciences, Huizhou 516000, Guangdong Province, China}}

\begin{document}
\title{\boldmath The heavy flavor conserving hadronic weak decay of the ground-state bottom baryons}

\author{Peng-Yu Niu}\email{niupy@m.scnu.edu.cn}
\affiliation{Guangdong Provincial Key Laboratory of Nuclear Science, Institute of Quantum Matter, South China Normal University, Guangzhou 510006, China}
\affiliation{Guangdong-Hong Kong Joint Laboratory of Quantum Matter, Southern Nuclear Science Computing Center, South China Normal University, Guangzhou 510006, China}

\author{Qian Wang}\email{qianwang@m.scnu.edu.cn}
\affiliation{Guangdong Provincial Key Laboratory of Nuclear Science, Institute of Quantum Matter, South China Normal University, Guangzhou 510006, China}
\affiliation{Guangdong-Hong Kong Joint Laboratory of Quantum Matter, Southern Nuclear Science Computing Center, South China Normal University, Guangzhou 510006, China}
\scnt

\author{Qiang Zhao}\email{zhaoq@ihep.ac.cn}
\affiliation{Institute of High Energy Physics, Chinese Academy of Sciences, Beijing 100049, China }
\affiliation{University of Chinese Academy of Sciences, Beijing 100049, China}
\affiliation{Center for High Energy Physics, Henan Academy of Sciences, Zhengzhou 450046, China}

\date{\today}

%%%%%%%%%%%%%%%%%%%%%%%%%%%%%%%%%%%%%%%%%%%%%%%%%%%%%%%%%%%%%%%%%%%%%%%%%%%%%%%%
\begin{abstract}
In this work the heavy flavor conserving (HFC) hadronic weak decays of bottom baryons are studied in the framework of the nonrelativistic constituent quark model (NRCQM). 
We show that the pole terms play an indispensable role in the description of the branching ratio of $\Xi_b^-\to \Lambda_b^0 \pi^-$. With the pole terms included we can make reliable predictions for $\Xi_b^0\to \Lambda_b^0 \pi^0$. A combined study of the HFC hadronic weak decays allows us to make a reasonable prediction for $\Omega_b^-\to\Xi_b^{-(0)}\pi^{0(-)}$, which can be searched for at LHCb and Belle-II experiments.

\end{abstract}
%\pacs{xx.Gp,  yy.Rj, zz.Dh}

\maketitle

%%%%%%%%%%%%%%%%%%%%%%%%%%%%%%%%%%%%%%%%%%%%%%%%%%%%%%%%%%%%%%%%%%%%%%%%%%%%%%%%
\section{Introduction}

For the hadronic weak decay of heavy flavor baryons, the heavy quark is usually involved in the weak interaction. However, there is a special hadronic weak decay process, i.e. the heavy-flavor-conserving (HFC) hadronic weak decay, where the heavy flavor is conserved, such as $\Xi_c \to \Lambda_c \pi$ in the charm sector, and $\Xi_b\to \Lambda_b^0\pi$,  $\Omega_b\to \Xi_b \pi$ in the bottom sector.
These processes, though limited by the phase space, have provided a unique way for probing the hadron structure and the transition mechanism in the non-perturbative QCD (non-pQCD) regime.

As pointed in Refs.~\cite{Voloshin:2000et, LHCb:2020gge, Niu:2021qcc, Cheng:2022jbr}, the charm quark will not only behave as a spectator, but also be involved in the weak interaction via $c s\to cd$. More specifically, when the pole terms induced by the $c s\to cd$ conversion are  almost on-shell, they will largely enhance the transition amplitude. 
For instance, the intermediate $\Sigma_c$ pole-term contribution is  the key to understanding the experimental data for $\Xi_c \to \Lambda_c \pi$~\cite{Niu:2021qcc}. In contrast, the bottom quark in the bottom baryon hadronic weak decays will behave only as a spectator. As the result, in the heavy quark limit, both the light degrees of freedom and total angular momentum should be conserved, individually. Thus, the $\Xi_b\to \Lambda_b^0\pi$ process is an $S$-wave decay, which is a parity-violating process, with the leading order transition operators in the heavy quark limit (more detailed discussion can be found in Sec.~\ref{sec:result}).

The interesting HFC hadronic weak decay of heavy baryons was noticed for the first time since 1990s and studied with various methods (see Refs.~\cite{Cheng:1992ff, Sinha:1999tc,Voloshin:2000et, Li:2014ada,Faller:2015oma, Gronau:2016xiq, Voloshin:2019ngb, Niu:2021qcc, Cheng:2022jbr,Cheng:2015ckx,Gronau:2015jgh, Cheng:2022kea} and references there in). Unfortunately, it is hard to simultaneously describe all the experimental data well. In our previous work~\cite{Niu:2021qcc}, we proposed that the pole terms are crucial for understanding the branching ratio of $\Xi_c^0\to \Lambda_c \pi^-$. Meanwhile, our calculation favors the lower bound $(0.57\pm 0.21)\%$ of the branching ratio $\mathrm{Br}(\Xi^-_b\to \Lambda_b^0\pi^-)$ measured by LHCb collaboration~\cite{LHCb:2015une}. However, the follow-up measurements indicate a larger branching ratio. 

The experimental measurement of the branching ratio largely depends on the ratio of the fragmentation fractions $f_{\Xi_b^-}$ and $f_{\Lambda_b^0}$, which reflects the fragmentation fractions of $b\to \Xi_b^-$ and $b\to \Lambda_b^0$, respectively. 
Based on the most earliest measurement~\cite{LHCb:2015une} 
\begin{align}
\frac{f_{\Xi_b^-}}{f_{\Lambda_b^0}}\mathrm{Br}(\Xi_b^-\to \Lambda_b^0 \pi^-)=(5.7\pm1.8^{+0.8}_{-0.9})\times 10^{-4},
\end{align}
and the assumption that $f_{\Xi_b^-}/f_{\Lambda_b^0}$ is bounded between 0.1 and 0.3, which is estimated with the measurement of the production rates of other strange particles relative to their nonstrange counterparts~\cite{LHCb:2015une}, the $\mathrm{Br}(\Xi^-\to \Lambda \pi^-)$ would lie in the range between $(0.57\pm 0.21)\%$ and $(0.19 \pm 0.07)\%$.

Recently, the relative rate 
\begin{align}
\frac{f_{\Xi_b^-}}{f_{\Lambda_b^0}}\mathrm{Br}(\Xi_b^-\to \Lambda_b^0 \pi^-)=(7.3\pm0.8\pm 0.6)\times 10^{-4},
\end{align}
was updated with the proton-proton collision data at center-of mass energy of $\sqrt{s}=13$ TeV with an integrated luminosity of 5.5 $\mathrm{fb}^{-1}$~\cite{LHCb:2023tma}. The center value is slightly larger than the old one. Benefiting from the independent measurement of $f_{\Xi_b^-}/f_{\Lambda_b^0}$~\cite{LHCb:2019sxa}, a more accurate of $\mathrm{Br}(\Xi_b^-\to \Lambda_b^0\pi^-)$ is determined to be:
\begin{align}
\mathrm{Br}(\Xi_b^-\to \Lambda_b^0\pi^-)= (0.89\pm 0.10\pm 0.07\pm 0.29)\%.
\end{align}
The last uncertainty is due to the SU(3) flavor symmetry assumption in the determination of $f_{\Xi_b^-}/f_{\Lambda_b^0}$. Compared with the previous measurement results, the uncertainty of the range of $\mathrm{Br}(\Xi_b^-\to \Lambda_b^0\pi^-)$ has significantly decreased. It provides a strict limit to the phenomenology study for the HFC hadronic weak decay of the bottom baryons.

In this work, we revisit the decay of $\Xi_b\to \Lambda_b^0\pi$ in the framework of the non-relativistic constituent quark model (NRCQM), which can still provide quantitative descriptions of the transitions, although the model uncertainties of the NRCQM are unavoidable. In order to have a better constaint on the NRCQM parameters, we note that the mass of the constituent quarks has a recognized range of values. Relatively large uncertainties may arise from the coupling constants of the Hamiltonian, for instance, the spring constant varies in different models or in different flavor systems. In our previous work~\cite{Niu:2021qcc}, a plausible range of the spring constant is taken, based on the studies of the baryon spectrum. Here, except for the mass of quarks, all the other parameter values are extracted from the experimental data of bottom baryon decay processes.
Considering the lack of experimental data for the HFC hadronic weak decays of the bottom baryons, the strong decays of $\Sigma_b$ are involved to restrict the values of the parameters. Meanwhile, all the parameters will be compared with the one extracted from the heavy baryon spectrum.

This paper is organized as follows. The transition diagrams of $\Xi_b \to \Lambda_b^0\pi$ and $\Omega_b \to \Xi_b \pi$ and a brief introduction of the NRQCM are presented in Sec.~\ref{sec:framework}. The results and discussions are given in Sec.~\ref{sec:result}. The last part is a brief summary.

%%%%%%%%%%%%%%%%%%%%%%%%%%%%%%%%%%%%%%%%%%%%%%%%%%%%%%%%%%%%%%%%%%%%%%%%%%%%%%%%
\section{Framework}
\label{sec:framework}
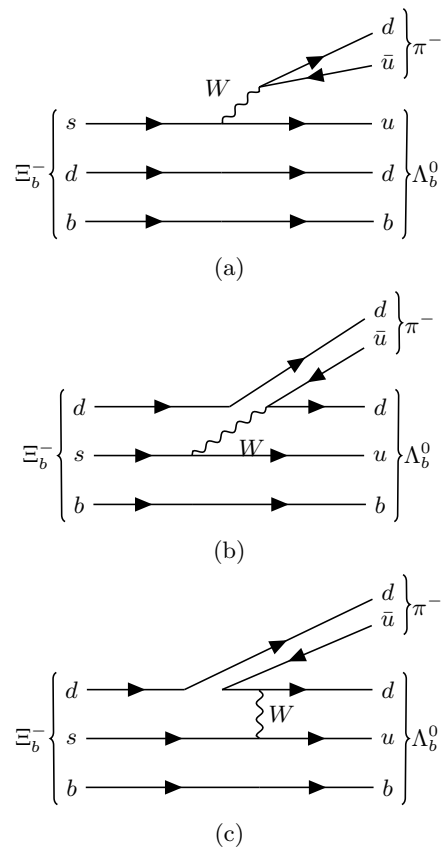
\begin{figure}[htbp]
\centering
\begin{subfigure}[h ]{0.45\textwidth}
\begin{tikzpicture}[line width=0.6pt]
\begin{feynman}
\vertex (a1) {$ s $};
\vertex[right=2.5cm of a1] (w1);
\vertex[right=2cm of a1] (a2);
\vertex[right=2cm of a2] (a3){$ u $};
\vertex[below=2em of a1] (b1) {$ d $};
\vertex[below=2em of a2] (b2) ;
\vertex[below=2em of a3] (b3) {$ d $};
\vertex[below=2em of b1] (c1) {$ b $};
\vertex[below=2em of b2] (c2);
\vertex[below=2em of b3] (c3) {$ b $};
\vertex[above=4em of a3]   (d2) {$d$};
\vertex[above=2.5em of a3] (d3) {$\bar u$};
\vertex[above=1.5em of w1] (e1);
\diagram* {
(a1) -- [fermion] (a2) -- [fermion] (a3),
(b1) -- [fermion] (b2) -- [fermion] (b3),
(c1) -- [fermion] (c2) -- [fermion] (c3),
(e1) -- [fermion] (d2),
(d3) -- [fermion] (e1),
(a2) -- [boson, edge label=$W$] (e1),
};
\draw [decoration={brace}, decorate] (c1.south west) -- (a1.north west)
node [pos=0.5, left] {$\Xi_b^{-}$};
\draw [decoration={brace}, decorate] (a3.north east) -- (c3.south east)
node [pos=0.5, right] {$\Lambda_b^0$};
\draw [decoration={brace}, decorate] (d2.north east) -- (d3.south east)
node [pos=0.5, right] {$\pi^-$};
\end{feynman}
\end{tikzpicture}
\caption{ }
\end{subfigure}
\\
\begin{subfigure}[htbp! ]{0.45\textwidth}
\begin{tikzpicture}[line width=0.6pt]
\begin{feynman}
\vertex (a1) {$ d $};
\vertex[right=2cm of a1] (a2);
\vertex[right=4cm of a1] (a3){$ d $};
\vertex[below=2em of a1] (b1) {$ s $};
\vertex[right=1.5 of b1] (b2) ;
\vertex[below=2em of a3] (b3) {$ u $};
\vertex[below=2em of b1] (c1) {$ b $};
\vertex[right=1.5cm of c1] (c2);
\vertex[below=2em of b3] (c3) {$ b $};
\vertex[above=4 em of a3]  (d1) {$d$};
\vertex[above=2.8em of a3] (d2) {$\bar u$};
\vertex[right=0.5cm of a2] (w1);
\vertex[right=1.5 of b1] (w2) ;
\diagram* {
(a1) -- [fermion] (a2),
(w1) -- [fermion] (a3),
(b1) -- [fermion] (w2) -- [fermion] (b3),
(c1) -- [fermion] (c2) -- [fermion] (c3),
(a2) -- [fermion] (d1),
(d2) -- [fermion] (w1),
(w1) -- [boson, edge label=$W$] (w2),
};
\draw [decoration={brace}, decorate] (c1.south west) -- (a1.north west)
node [pos=0.5, left] {$\Xi_b^{-}$};
\draw [decoration={brace}, decorate] (a3.north east) -- (c3.south east)
node [pos=0.5, right] {$\Lambda_b^0$};
\draw [decoration={brace}, decorate] (d1.north east) -- (d2.south east)
node [pos=0.5, right] {$\pi^-$};
\end{feynman}
\end{tikzpicture}
\caption{}
\end{subfigure}
~
\begin{subfigure}[htbp! ]{0.45\textwidth}
\begin{tikzpicture}[line width=0.6pt]
\begin{feynman}
\vertex (a1) {$ d $};
\vertex[right=1.5cm of a1] (a2);
\vertex[right=0.5cm of a2] (a3);
\vertex[right=0.5cm of a3] (a4);
\vertex[right=1.5cm of a4] (a5){$ d $};
\vertex[below=2em of a1] (b1) {$ s $};
\vertex[below=2em of a4] (b2) ;
\vertex[below=2em of a5] (b3) {$ u $};
\vertex[below=2em of b1] (c1) {$ b $};
\vertex[below=2em of b2] (c2);
\vertex[below=2em of b3] (c3) {$ b $};
\vertex[above=4em of a5]   (d1) {$d$};
\vertex[above=2.9em of a5] (d2) {$\bar u$};
\vertex[right=2.5cm of a1] (w1);
\vertex[below=2em of w1] (w2);
\diagram* {
(a1) -- [fermion] (a2),
(a3) -- [fermion] (a5),
(b1) -- [fermion] (b2) -- [fermion] (b3),
(c1) -- [fermion] (c2) -- [fermion] (c3),
(a2) -- [fermion] (d1),
(d2) -- [fermion] (a3),
(w1) -- [boson, edge label=$W$] (w2),
};
\draw [decoration={brace}, decorate] (c1.south west) -- (a1.north west)
node [pos=0.5, left] {$\Xi_b^{-}$};
\draw [decoration={brace}, decorate] (a5.north east) -- (c3.south east)
node [pos=0.5, right] {$\Lambda_b^0$};
\draw [decoration={brace}, decorate] (d1.north east) -- (d2.south east)
node [pos=0.5, right] {$\pi^-$};
\end{feynman}
\end{tikzpicture}
\caption{}
\end{subfigure}
\caption{The transition diagrams of the $\Xi_b^{-} \to  \Lambda_b^0~ \pi^-$. (a) DPE process; (b) CS process; (c) pole term.}
\label{fig:Xibm}
\end{figure}

\begin{figure}[htbp]
\centering
\begin{subfigure}[h ]{0.45\textwidth}
\begin{tikzpicture}[line width=0.6pt]
\begin{feynman}
\vertex (a1) {$ s $};
\vertex[right=1.8cm of a1] (a2);
\vertex[right=2.5cm of a1] (a3);
\vertex[right=4cm of a1] (a4){$ d $};
\vertex[below=2em of a1] (b1) {$ s $};
\vertex[below=2em of a2] (b2) ;
\vertex[below=2em of a4] (b3) {$ s $};
\vertex[below=2em of b1] (c1) {$ b $};
\vertex[below=2em of b2] (c2);
\vertex[below=2em of b3] (c3) {$ b $};
\vertex[above=4.4em of a4] (d1) {$u$};
\vertex[above=3em of a4] (d2) {$\bar u$};
\diagram* {
(a1) -- [fermion] (a2),
(a3) -- [fermion] (a4),
(b1) -- [fermion] (b3),
(c1) -- [fermion] (c3),
(a3) -- [boson, edge label=$W$] (a2),
(a2) -- [fermion] (d1),
(d2) -- [fermion] (a3),
};
\draw [decoration={brace}, decorate] (c1.south west) -- (a1.north west)
node [pos=0.5, left] {$\Omega_b^-$};
\draw [decoration={brace}, decorate] (a4.north east) -- (c3.south east)
node [pos=0.5, right] {$\Xi_b^-$};
\draw [decoration={brace}, decorate] (d1.north east) -- (d2.south east)
node [pos=0.5, right] {$\pi^0$};
\end{feynman}
\end{tikzpicture}
\end{subfigure}
\caption{The transition diagram of the $\Omega_b^-\to  \Xi_b^- ~ \pi^0$. Only CS process is allowed.}
\label{fig:Omegapiz}
\end{figure}
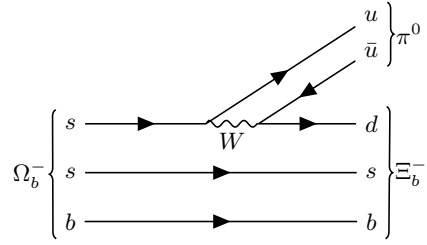

The transition diagrams for the $\Xi_b \to \Lambda_b^0\pi$ and $\Omega_b \to \Xi_b\pi$ processes are illustrated by Figs.~\ref{fig:Xibm}-~\ref{fig:Omegapiz}. The hadronic weak decay of bottom baryon involves three kinds of processes. The first one is the direct pion emission (DPE) as shown by Fig.~\ref{fig:Xibm}(a). The second one is the color-suppressed(CS) process that can be divided into two types.
The first type describes the transition that the internal emitted $W$ produces a new pair of quark and antiquark, which will immediately combine with one of the quarks of the initial baryons to form the final meson. We label it as type one of the color-suppressed process (CS-I) (Fig.~\ref{fig:Xibm}(b)), while the second type of color-suppressed process (CS-II) describes the transition that the constituent quarks of the final meson are all from the  weak interaction vertex as shown by Fig.~\ref{fig:Xibz}(b). The third typical transition is the pole term which can also be divided into two types. The first type is labeled as SW, where the meson is first produced by the strong interaction and then an intermediate state will propagate until two of the constituent quarks interact and convert into the final state hadron via the $W$ exchange as shown by Fig.~\ref{fig:Xibm}(c). The second type of the pole term,  labeled as WS, describes the transition where the time order of the strong interaction and the weak interaction is reversed, as shown by Fig.~\ref{fig:Xibz}(c).
It should be stressed that the two transitions via Fig.~\ref{fig:Xibm}(b) and (c) have totally different analytical structures and are dynamically different. To be more specific, Fig.~\ref{fig:Xibm}(b) only contains the weak interaction and does not contain pole structures in its transition amplitude. In contrast, in Fig.~\ref{fig:Xibm}(c) the strong and the weak interaction are involved non-locally. The transition  proceeds via two steps and has a pole structure in the transition amplitude that can manifests itself via experimental observables. This feature also makes it necessary to explicitly consider the pole terms in some of these weak decay transitions. As demonstrated in Ref.~\cite{Niu:2021qcc,Cheng:2022jbr} the pole terms play an indispensable role in the HFC hadronic weak decays. As follows, we will present the wavefunctions and transition operators adopted in the calculation of the decay amplitudes.

%%%%%%%%%%%%%%%%%%%%%%%%%%%%%%%%%%%%%%%%
\begin{figure}[htbp]
\centering
\begin{subfigure}[htbp! ]{0.45\textwidth}
\begin{tikzpicture}[line width=0.6pt]
\begin{feynman}
\vertex (a1) {$ u $};
\vertex[right=2cm of a1] (a2);
\vertex[right=4cm of a1] (a3){$ d $};
\vertex[below=2em of a1] (b1) {$ s $};
\vertex[right=1.5 of b1] (b2) ;
\vertex[below=2em of a3] (b3) {$ u $};
\vertex[below=2em of b1] (c1) {$ b $};
\vertex[right=1.5cm of c1] (c2);
\vertex[below=2em of b3] (c3) {$ b $};
\vertex[above=4 em of a3]  (d1) {$u$};
\vertex[above=2.8em of a3] (d2) {$\bar u$};
\vertex[right=0.5cm of a2] (w1);
\vertex[right=1.5 of b1] (w2) ;
\diagram* {
(a1) -- [fermion] (a2),
(w1) -- [fermion] (a3),
(b1) -- [fermion] (w2) -- [fermion] (b3),
(c1) -- [fermion] (c2) -- [fermion] (c3),
(a2) -- [fermion] (d1),
(d2) -- [fermion] (w1),
(w1) -- [boson, edge label=$W$] (w2),
};
\draw [decoration={brace}, decorate] (c1.south west) -- (a1.north west)
node [pos=0.5, left] {$\Xi_b^{0}$};
\draw [decoration={brace}, decorate] (a3.north east) -- (c3.south east)
node [pos=0.5, right] {$\Lambda_b^0$};
\draw [decoration={brace}, decorate] (d1.north east) -- (d2.south east)
node [pos=0.5, right] {$\pi^0$};
\end{feynman}
\end{tikzpicture}
\caption{}
\end{subfigure}
~
\begin{subfigure}[h ]{0.45\textwidth}
\begin{tikzpicture}[line width=0.6pt]
\begin{feynman}
\vertex (a1) {$ s $};
\vertex[right=1.8cm of a1] (a2);
\vertex[right=2.5cm of a1] (a3);
\vertex[right=4cm of a1] (a4){$ d $};
\vertex[below=2em of a1] (b1) {$ u $};
\vertex[below=2em of a2] (b2) ;
\vertex[below=2em of a4] (b3) {$ u $};
\vertex[below=2em of b1] (c1) {$ b $};
\vertex[below=2em of b2] (c2);
\vertex[below=2em of b3] (c3) {$ b $};
\vertex[above=4.4em of a4] (d1) {$u$};
\vertex[above=3em of a4] (d2) {$\bar u$};
\diagram* {
(a1) -- [fermion] (a2),
(a3) -- [fermion] (a4),
(b1) -- [fermion] (b3),
(c1) -- [fermion] (c3),
(a3) -- [boson, edge label=$W$] (a2),
(a2) -- [fermion] (d1),
(d2) -- [fermion] (a3),
};
\draw [decoration={brace}, decorate] (c1.south west) -- (a1.north west)
node [pos=0.5, left] {$\Xi_b^{0}$};
\draw [decoration={brace}, decorate] (a4.north east) -- (c3.south east)
node [pos=0.5, right] {$\Lambda_b^0$};
\draw [decoration={brace}, decorate] (d1.north east) -- (d2.south east)
node [pos=0.5, right] {$\pi^0$};
\end{feynman}
\end{tikzpicture}
\caption{}
\end{subfigure}
~
\begin{subfigure}[htbp! ]{0.45\textwidth}
\begin{tikzpicture}[line width=0.6pt]
\begin{feynman}
\vertex (a1) {$ u $};
\vertex[right=1.5cm of a1] (a2);
\vertex[right=0.5cm of a2] (a3);
\vertex[right=0.5cm of a3] (a4);
\vertex[right=1.5cm of a4] (a5){$ d $};
\vertex[below=2em of a1] (b1) {$ s $};
\vertex[below=2em of a2] (b2) ;
\vertex[below=2em of a5] (b3) {$ u $};
\vertex[below=2em of b1] (c1) {$ b $};
\vertex[below=2em of b2] (c2);
\vertex[below=2em of b3] (c3) {$ b $};
\vertex[above=4em of a5]   (d1) {$d$};
\vertex[above=2.8em of a5] (d2) {$\bar d$};
\diagram* {
(a1) -- [fermion] (a3),
(a4) -- [fermion] (a5),
(b1) -- [fermion] (b2) -- [fermion] (b3),
(c1) -- [fermion] (c2) -- [fermion] (c3),
(a3) -- [fermion] (d1),
(d2) -- [fermion] (a4),
(a2) -- [boson, edge label=$W$] (b2),
};
\draw [decoration={brace}, decorate] (c1.south west) -- (a1.north west)
node [pos=0.5, left] {$\Xi_b^{0}$};
\draw [decoration={brace}, decorate] (a5.north east) -- (c3.south east)
node [pos=0.5, right] {$\Lambda_b^0$};
\draw [decoration={brace}, decorate] (d1.north east) -- (d2.south east)
node [pos=0.5, right] {$\pi^0$};
\end{feynman}
\end{tikzpicture}
\caption{}
\end{subfigure} 
\\
\begin{subfigure}[htbp! ]{0.45\textwidth}
\begin{tikzpicture}[line width=0.6pt]
\begin{feynman}
\vertex (a1) {$ s $};
\vertex[right=1.5cm of a1] (a2);
\vertex[right=0.5cm of a2] (a3);
\vertex[right=0.5cm of a3] (a4);
\vertex[right=1.5cm of a4] (a5){$ u $};
\vertex[below=2em of a1] (b1) {$ u $};
\vertex[below=2em of a2] (b2) ;
\vertex[below=2em of a5] (b3) {$ d $};
\vertex[below=2em of b1] (c1) {$ b $};
\vertex[below=2em of b2] (c2);
\vertex[below=2em of b3] (c3) {$ b $};
\vertex[above=4em of a5]   (d1) {$u$};
\vertex[above=2.8em of a5] (d2) {$\bar u$};
\diagram* {
(a1) -- [fermion] (a3),
(a4) -- [fermion] (a5),
(b1) -- [fermion] (b2) -- [fermion] (b3),
(c1) -- [fermion] (c2) -- [fermion] (c3),
(a3) -- [fermion] (d1),
(d2) -- [fermion] (a4),
(a2) -- [boson, edge label=$W$] (b2),
};
\draw [decoration={brace}, decorate] (c1.south west) -- (a1.north west)
node [pos=0.5, left] {$\Xi_b^{0}$};
\draw [decoration={brace}, decorate] (a5.north east) -- (c3.south east)
node [pos=0.5, right] {$\Lambda_b^0$};
\draw [decoration={brace}, decorate] (d1.north east) -- (d2.south east)
node [pos=0.5, right] {$\pi^0$};
\end{feynman}
\end{tikzpicture}
\caption{}
\end{subfigure} 
~
\begin{subfigure}[htbp! ]{0.45\textwidth}
\begin{tikzpicture}[line width=0.6pt]
\begin{feynman}
\vertex (a1) {$ u $};
\vertex[right=1.5cm of a1] (a2);
\vertex[right=0.5cm of a2] (a3);
\vertex[right=0.5cm of a3] (a4);
\vertex[right=1.5cm of a4] (a5){$ d $};
\vertex[below=2em of a1] (b1) {$ s $};
\vertex[below=2em of a4] (b2) ;
\vertex[below=2em of a5] (b3) {$ u $};
\vertex[below=2em of b1] (c1) {$ b $};
\vertex[below=2em of b2] (c2);
\vertex[below=2em of b3] (c3) {$ b $};
\vertex[above=4em of a5]   (d1) {$u$};
\vertex[above=2.9em of a5] (d2) {$\bar u$};
\vertex[right=2.5cm of a1] (w1);
\vertex[below=2em of w1] (w2);
\diagram* {
(a1) -- [fermion] (a2),
(a3) -- [fermion] (a5),
(b1) -- [fermion] (b2) -- [fermion] (b3),
(c1) -- [fermion] (c2) -- [fermion] (c3),
(a2) -- [fermion] (d1),
(d2) -- [fermion] (a3),
(w1) -- [boson, edge label=$W$] (w2),
};
\draw [decoration={brace}, decorate] (c1.south west) -- (a1.north west)
node [pos=0.5, left] {$\Xi_b^{0}$};
\draw [decoration={brace}, decorate] (a5.north east) -- (c3.south east)
node [pos=0.5, right] {$\Lambda_b^0$};
\draw [decoration={brace}, decorate] (d1.north east) -- (d2.south east)
node [pos=0.5, right] {$\pi^0$};
\end{feynman}
\end{tikzpicture}
\caption{}
\end{subfigure}
\caption{The transition diagrams of the $\Xi_b^{0}\to  \Lambda_b^0 \pi^0$. (a) CS process; (b)-(d) pole terms. }
\label{fig:Xibz}
\end{figure}
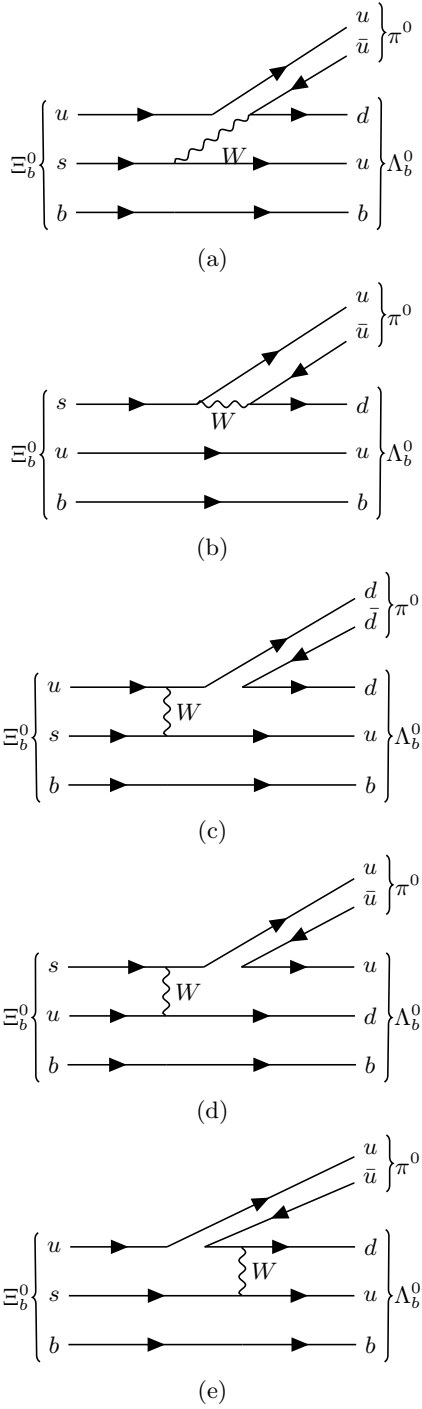

%%%%%%%%%%%%%%%%%%%%%%%%%%%%%%%
\begin{figure}[htbp]
\centering
\begin{subfigure}[h ]{0.45\textwidth}
\begin{tikzpicture}[line width=0.6pt]
\begin{feynman}
\vertex (a1) {$ s $};
\vertex[right=2.5cm of a1] (w1);
\vertex[right=2cm of a1] (a2);
\vertex[right=2cm of a2] (a3){$ u $};
\vertex[below=2em of a1] (b1) {$ s $};
\vertex[below=2em of a2] (b2) ;
\vertex[below=2em of a3] (b3) {$ s $};
\vertex[below=2em of b1] (c1) {$ b $};
\vertex[below=2em of b2] (c2);
\vertex[below=2em of b3] (c3) {$ b $};
\vertex[above=4em of a3]   (d2) {$d$};
\vertex[above=2.5em of a3] (d3) {$\bar u$};
\vertex[above=1.5em of w1] (e1);
\diagram* {
(a1) -- [fermion] (a2) -- [fermion] (a3),
(b1) -- [fermion] (b2) -- [fermion] (b3),
(c1) -- [fermion] (c2) -- [fermion] (c3),
(e1) -- [fermion] (d2),
(d3) -- [fermion] (e1),
(a2) -- [boson, edge label=$W$] (e1),
};
\draw [decoration={brace}, decorate] (c1.south west) -- (a1.north west)
node [pos=0.5, left] {$\Omega_b^-$};
\draw [decoration={brace}, decorate] (a3.north east) -- (c3.south east)
node [pos=0.5, right] {$\Xi_b^0$};
\draw [decoration={brace}, decorate] (d2.north east) -- (d3.south east)
node [pos=0.5, right] {$\pi^-$};
\end{feynman}
\end{tikzpicture}
\end{subfigure}
\caption{The transition diagrams of the $\Omega_b^-\to  \Xi_b^0~ \pi^-$. Only DPE process is allowed.}
\label{fig:Omegapim}
\end{figure}
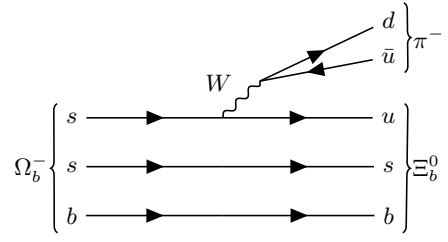 

%%%%%%%%%%%%%%%%%%%%%%%%%%%%

%%%%%%%%%%%%%%%%%%%%%%%%%%%%%%%%%%%%%%%%%%%%%%%%%%%%%%%%%%%%%%%%%%%%%%%%%%%%%%%%
\subsection{The wave function of hadrons}
\label{app:wavefunction}

In the framework of NRCQM, the hadron wave function is constructed with the color, flavor, spin and spatial wave functions. Apart from the color wave function, the standard $\mathrm {SU(6)} \otimes \mathrm{O}(3)$ wave functions~\cite{LeYaouanc:1988fx} are used in this work. The total wavefunctions for the bottom baryons and the pseudoscalar meson can be find in Ref.~\cite{Niu:2021qcc}. For the wave function of the quark model, the main uncertainty stems from the spatial wave function. Technically, it is limited by  the effective Hamiltonian for the quark and/or antiquark systems which is insufficient for the description of the nonperturbative QCD.

Generally, the effective Hamiltonian of the quark systems includes the kinetic energy, confinement potential, and short-range interaction or hyperfine interaction. Ignoring the hyperfine interaction and taking the harmonic oscillator approximation, the effective Hamiltonian for the three-quark system \cite{Isgur:1978xj, LeYaouanc:1988fx} can be written as
\begin{align}
\label{eq:HOH}
H=\sum_{i=1}^3 \frac{\bm p_i^2}{2m_i^2}+\sum_{i\neq j}\frac{1}{2}K|\bm r_i-\bm r_j|^2 \ ,
\end{align}
where $\bm r_i$ and $\bm p_i$ are the coordinate and momentum for the $i$th quark. $K$ is the spring constant that describes the strength of the interaction between quarks.

With the coordinate definitions
\begin{align}
\bm \rho&\equiv\frac{1}{\sqrt 2} (\bm r_1-\bm r_2),\\
\bm \lambda&\equiv\sqrt{\frac{2}{3}}\left( \frac{m_1}{m_1+m_2} \bm r_1+\frac{m_2}{m_1+m_2} \bm r_2- \bm r_3 \right),
\end{align}
the corresponding momenta can be obtained as
\begin{align}
\bm p_\rho    &=m_r \dot{\bm \rho}=\sqrt{2}\left( \frac{m_2}{m_1+m_2}  \bm p_1- \frac{m_1}{m_1+m_2}  \bm p_2\right), \\
\bm p_\lambda &=m_\lambda \dot{\bm \lambda}= \sqrt{\frac{3}{2}}\left( \frac{m_3}{M}  \bm p_1+\frac{m_3}{M}  \bm p_2-\frac{m_1+m_2}{M}  \bm p_3 \right).
\end{align}
As the result, the Hamiltonian can be expressed as
\begin{align}
H=\frac{\bm{P}^2}{2 M}+\frac{\bm{p}_\rho^2}{2 m_\rho}+\frac{\bm{p}_\lambda^2}{2 m_\lambda}+\frac{1}{2} m_\rho \omega_\rho^2 \bm{\rho}^2+\frac{1}{2} m_\lambda \omega_\lambda^2 \bm\lambda^2 \ ,
\end{align}
where $\bm P$ is the total momentum of the baryon and $M=m_1+m_2+m_3$ is the total mass; $m_\rho$ and $m_\lambda$ are the reduce mass and defined as
\begin{align}
m_\rho= 2\frac{m_1m_2}{m_1+m_2},~m_\lambda=\frac{3}{2}\frac{m_3(m_1+m_2)}{M}.
\end{align}

Then, the spatial wave function, that is also referred to as the harmonic oscillator wave function, can be easily obtained and reads as
\begin{align}
\Psi_{N L L_z}(\bm P,\bm p_\rho,\bm p_\lambda)
&=\delta^3(\bm P-\bm P_c)\sum_m \langle l_\rho,m;l_\lambda,L_z-m|L,L_z \rangle
\notag \\
&\times \psi^{\alpha_\rho}_{n_\rho l_\rho m }(\bm p_\rho)  
\psi^{\alpha_\lambda}_{n_\lambda l_\lambda L_z-m }(\bm p_\lambda),
\end{align}
where
\begin{align}
\psi^\alpha_{nlm}(\bm p)&=(i)^l(-1)^n \left[\frac{2n!}{(n+l+1/2)!} \right]^{\frac12} \frac{1}{\alpha^{l+3/2}}\mathcal{Y}_{lm}(\bm p) \notag \\
&\times L_n^{l+1/2}(\bm p^2/\alpha^2)\exp\left({-\frac{\bm p^2}{2\alpha^2}}\right) \ .
\end{align}
$\alpha_\rho$ and $\alpha_\lambda$ are the harmonic oscillator strengths that are determined by the reduce mass $m_{\rho/\lambda}$ and the frequencies $\omega_{\rho/\lambda}$:
\begin{align}
\label{eq:arholambda}
\alpha_\rho^2=m_\rho\omega_\rho,~\alpha_\lambda^2=m_\lambda\omega_\lambda ,
\end{align}
with
\begin{align}
\omega_\rho^2=\frac{3K}{m_\rho},~\omega_\lambda^2=\frac{3K}{m_\lambda}.
\end{align}

When the hyperfine interaction considered, the spatial wave function of the baryons can be expanded with the harmonic oscillator wave functions. Here, we still use the harmonic oscillator wave functions determined by Eq.~(\ref{eq:HOH}). But the values of the harmonic oscillator strengths $\alpha_{\rho/\lambda}$ should be different from the ones that appear in the realistic spatial wave function. It can be understood that part of the effects from the hyperfine interaction are absorbed into $\alpha_{\rho/\lambda}$.
%that can also be considered as the parameters of the wave function instead of $K$.

For the light mesons as a quark-antiquark system, the spatial wave function of the ground state is expressed as:
\begin{equation}
\Psi(\bm p_1, \bm p_2)= \frac{1}{\pi^{3/4} R^{3/2}}\exp\left[-\frac{(\bm p_1-\bm p_2)^2}{8 R^2}\right],
\end{equation}
where $R$ characterizes the size of mesons and is determined by parameter $K$ and the quark masses. For the pion we take $R=0.28$ GeV that is the same as in Ref.~\cite{Niu:2021qcc}.

\subsection{The transition operators}

\subsubsection{The non-relativistic form of the weak interaction}

The effective Hamiltonian that describes the four-fermion interaction is written as ~\cite{LeYaouanc:1978ef, LeYaouanc:1988fx}:
\begin{align}
H_W=\frac{G_F}{\sqrt 2}\int d \bm x \frac12 \{J^{-,\mu}(\bm x),J^{+}_{\mu}(\bm x) \},
\end{align}
where 
\begin{align}
J^{+,\mu}(\bm x)&=
\begin{pmatrix}\bar u&\bar c \end{pmatrix}
\gamma^\mu(1-\gamma_5)
\begin{pmatrix}\cos \theta_C & \sin \theta_C \\ -\sin \theta_C &\cos \theta_C \end{pmatrix} \begin{pmatrix} d\\s \end{pmatrix}, 
\end{align}
\begin{align}
J^{-,\mu}(\bm x)&=
\begin{pmatrix}\bar d &\bar s \end{pmatrix} \begin{pmatrix}\cos \theta_C & -\sin \theta_C \\ \sin \theta_C &\cos \theta_C \end{pmatrix} \gamma^\mu(1-\gamma_5)
\begin{pmatrix} u\\c \end{pmatrix} \ ,
\end{align}
where $\theta_C$ is the Cabibbo angle. The nonrelativistic form of the weak transition operators used in this work can be found in Ref.~\cite{ Niu:2021qcc}.
In this work, contributions from the QCD corrections will be absorbed into the quark model parameters, and higher order processes such as  the penguin diagrams~\cite{Buchalla:1995vs} are neglected since their contributions are highly suppressed.

For the processes shown by Fig.~\ref{fig:Xibm}(b), Fig.~\ref{fig:Xibz}(b) and Fig.~\ref{fig:Omegapiz}, the color index of the $d$ quark created by the weak interaction vertex is not arbitrary since the color indices of the other two quarks of the final baryon are fixed. Hence, these processes are color suppressed and share the same color factor of  $1/3$.

There exist the intermediate states in the pole terms. Only the contributions of ground states whose $J^p=1/2^+$ and the first excited bottom baryons whose $J^p=1/2^-$ are considered. All the possible states and their masses from different models are listed in Tab.~\ref{tab:cbmass}. The values of masses used in our calculation are listed in the last line.

\begin{table*}[htbp]
\centering
\caption{The masses of the bottom baryons (in unit of MeV). Only their central values are listed. Here, we use the symbol $|^{(2s+1)}l_\mathrm{sym}\rangle$ to label the baryon state. $l$ and $s$ are the total orbital angular momentum and  the total spin, respectively. The subscript `$\mathrm{sym}$' is used to label the symmetry property of the corresponding  spatial wavefunctions. The $\lambda$ and $\rho$ mode denote that the spatial wave function is symmetric or asymmetric under the interchange of the coordinate index of the first two quarks, respectively; `$\cdots$' means that the data are unavailable from the corresponding references. }
\begin{ruledtabular}
\begin{tabular}{l|llll|llll|llll|llll}
\multicolumn{1}{c|}{\multirow{2}[0]{*}{States}}& \multicolumn{4}{l|}{$\Xi_b$} & \multicolumn{4}{l|}{$\Xi'_b$} & \multicolumn{4}{l|}{$\Sigma_b$} & \multicolumn{4}{l}{$\Lambda_b$} \\ 
&$|^2 S\rangle$ &$|^2 P_\lambda\rangle$ &$|^2 P_\rho\rangle$ &$|^4 P_\rho\rangle$
&$|^2 S\rangle$ &$|^2 P_\lambda\rangle$ &$|^2 P_\rho\rangle$ &$|^4 P_\lambda\rangle$ 
&$|^2 S\rangle$ &$|^2 P_\lambda\rangle$ &$|^2 P_\rho\rangle$ &$|^4 P_\lambda\rangle$ 
&$|^2 S\rangle$ &$|^2 P_\lambda\rangle$ &$|^2 P_\rho\rangle$ &$|^4 P_\rho\rangle$ \\
\hline
PDG~\cite{ParticleDataGroup:2022pth} 
& 5792 &$\cdots$&$\cdots$&$\cdots$  
& 5935 &$\cdots$&$\cdots$&$\cdots$
& 5813 &$\cdots$&$\cdots$&$\cdots$
& 5612 &$\cdots$&$\cdots$&$\cdots$ \\
Ref.~\cite{Kim:2024tbf}
&5796   &6069 &5946-6589 &6054      
&5934   &6164 &$\cdots$  &6183
&5810   &6043 &$\cdots$  &6065       
&5620   &5914 &6207      &6233 \\
Ref.~\cite{Li:2024zze}
&$\cdots$    &$\cdots$   &$\cdots$   &$\cdots$     
&5943   &6223  &$\cdots$   &6225
&5820   &6100  &$\cdots$   &6089       
&$\cdots$    &$\cdots$   &$\cdots$   &$\cdots$ \\
Ref.~\cite{Bijker:2023zhh}     
&5812   &6077  &6243  &6285
&5935   &6201  &6366  &6216 
&$\cdots$  &$\cdots$ &$\cdots$ &$\cdots$       
&$\cdots$  &$\cdots$ &$\cdots$ &$\cdots$ \\
Ref.~\cite{Bijker:2020tns}  
&5784   &6048  &6214  &6226
&5926   &6190  &6356  &6203 
&$\cdots$  &$\cdots$ &$\cdots$ &$\cdots$       
&$\cdots$  &$\cdots$ &$\cdots$ &$\cdots$ \\
Ref.~\cite{Ortiz-Pacheco:2023kjn}  
&5812   &6077  &6243   &6258    
&5935   &6201  &6336  &6216
&5810   &6108  &6300  &6123       
&5615   &5913  &6106  &6121 \\
Ref.~\cite{Garcia-Tecocoatzi:2023btk}\footnote{Only the center values of three-quark predicted masses are listed.} 
&5806   &6079  &6248  &6271    
&5925   &6198  &6367  &6220
&5804   &6108  &6304  &6131       
&5613   &5918  &6114  &6137 \\
Ref.~\cite{Niu:2021qcc}
&5792   &6100   &6230   &6270 
&5935   &6220   &6350   &6250 
&5816   &6097   &6246   &6135 
&5620   &5912   &6236   &6273 \\
Used 
&5792   &6080   &6245   &6260
&5935   &6200   &6360   &6216
&5813   &6108   &6300   &6130
&5612   &5915   &6100   &6130
\\
\end{tabular}%
\end{ruledtabular}
\label{tab:cbmass}%
\end{table*}

\subsubsection{The non-relativistic form of the quark-meson interaction}

We employ the chiral quark model~\cite{Manohar:1983md, Li:1997gd,Zhao:2002id, Zhong:2007gp} to describe the strong transition vertex. In this model, the pion couples to the light quark through a chiral Lagrangian for the pseudoscalar meson coupling to the constituent quark. At the leading order, the effective Hamiltonian reads as~\cite{Manohar:1983md, Li:1997gd,Zhao:2002id,Zhong:2007gp}:
\begin{align}
\label{equ:Hpi}
H_{\chi}=\int d \bm x \frac{g_A^q}{f_m}\bar q(\bm x) \gamma_\mu \gamma_5 q(\bm x) \partial^\mu \phi(\bm x) ,
\end{align}
where $q(\bm x)$ is the light quark field. $\phi$ is the pseudoscalar octet and expressed as 
\begin{align}
\phi=\left(\begin{array}{ccc}
\frac{1}{\sqrt 2}\pi^0+\frac{1}{\sqrt 6}\eta & \pi^+ & K^+ \\
\pi^- & -\frac{1}{\sqrt 2}\pi^0+\frac{1}{\sqrt 6}\eta  & K^0\\
K^-   &\bar K^0  & -\sqrt{\frac{2}{3}}\eta
\end{array}\right) \ ,
\end{align}
where $g_A^q$ is the quark axial vector coupling constant that $g_A^q=1$~\cite{Arifi:2021orx}; $f_m$ is the pseudoscalar meson decay constant, for pion $f_\pi=132$ MeV. 

The non-relativistic form of $H_\chi$ is:
\begin{align}
\label{eq:Hdelta}
H_{\chi}&=\frac{\delta}{\sqrt{(2\pi)^3 2\omega_m}} \sum_{j=1}^2 \frac{1}{f_m}\left[\omega_m\left(\frac{\bm\sigma\cdot \bm p^j_f}{2m_f}+\frac{\bm\sigma\cdot \bm p^j_i}{2m_i}\right) -\bm \sigma \cdot \bm k\right] \notag \\
&\times \hat I^j_\chi \delta^3(\bm p^j_f+\bm k-\bm p_i^j) \ ,
\end{align}
where $\omega_m$ and $\bm k$ are the energy and momentum of the pseudoscalar meson, respectively; $\bm p^j_i$ and $\bm p^j_f$ are the initial and final momentum of the $j$th light quark, respectively; $\hat I^j_\chi$ is the isospin operator acting on the $j$th light quark and is written as:
\begin{align}
\hat I^j_\chi=\begin{cases}
a^\dagger_u a_d,                      &\mathrm{for}~ \pi^-,\\
a^\dagger_d a_u,                      &\mathrm{for}~ \pi^+,\\
\frac{1}{\sqrt2}(a^\dagger_u a_u-a^\dagger_d a_d),                      &\mathrm{for}~ \pi^0 \ ,
\end{cases}
\end{align}
where $a^\dagger_q$ and $a_q$ are the creation and annihilation operators for the $q$ quark, respectively.

In Eq.~(\ref{eq:Hdelta}), a dimensionless parameter $\delta$ is introduced to take into account uncertainties arising from the model~\cite{Zhong:2007gp}.
In our previous work~\cite{Niu:2021qcc}, $f_\pi$ is taken as $93$ MeV. This is equivalent to taking $\delta=\sqrt 2$. Different values for $\delta$ can be found in the literature. They may be due to the different normalization conditions introduced to the interacting vertex~\cite{Zhong:2007gp}.  %in the study of heavy hadrons strong decay, the $\delta$ is about $0.5$~\cite{Zhong:2007gp}.

%\subsection{The decay width}
With the wave functions and transition operators, the decay width can be expressed as
\begin{align}
\label{eq:decaywidth}
\Gamma(A\to B+C)&=8\pi^2\frac{|\bm k|E_B E_C}{M_A}\frac{1}{2 J_A+1} \notag \\
&\times\sum_\mathrm{spin} \left(|\mathcal M_{\mathrm{PC}}|^2+ |\mathcal M_{\mathrm{PV}}|^2 \right)
\end{align}
with the normalization convention 
\begin{align}
\langle B_f(\bm P_f) |B_i(\bm P_i) \rangle =\delta^3(\bm P_f- \bm P_i),
\end{align}
and $J_A$ is the spin of the initial state.

%%%%%%%%%%%%%%%%%%%%%%%%%%%%%%%%%%%%%%%%%%%%%%%%%%%%%%%%%%%%%%%%%%%%%%%%%%%%%%%%
\section{Numerical Results and discussion}
\label{sec:result}

\subsection{Parameters}

Within the harmonic oscillator approximation, i.e. neglecting the hyperfine effects~\cite{Isgur:1977ef,Isgur:1978xj,LeYaouanc:1988fx,Yoshida:2015tia}, there are two kind of model parameters in the Hamiltonian that are the quark mass and the spring constant $K$.
The masses of the quarks are taken as $m_u=m_d=0.3 ~\mathrm{GeV}, ~m_s=0.5~\mathrm{GeV}$, and $m_b=5~\mathrm{GeV}$ which are the same as our previous work~\cite{Niu:2021qcc}.

The spring constant $K$ is an essential parameter of the non-relativistic component quark model. Its value indicates the strengths of interactions between quarks. Both $K$ and the mass of the quarks determine the harmonic oscillator strengths, subsequently determining the root-mean-square radius of the hadrons, as shown in Sec.~\ref{sec:framework}. Ideally, $K$ is flavor independent, while, for the phenomenological approach, $K$ could take different values in different hadron systems. Usually, the value of $K$ is extracted by fitting the spectrum of the hadrons. For the bottom heavy baryons, the values of $K$ adopted in different works are listed in Tab.~\ref{tab:kValue} as a  comparison.

\begin{table}[htbp]
\centering
\caption{The ranges of the $K$ for the bottom baryons. In Ref.~\cite{Bijker:2020tns}, the value of $\alpha_\rho$ used in the calculation of the strong and electromagnetic decay is $\alpha_\rho=403$ MeV, and the $K$ is estimated with $\alpha_\rho^4=3Km_\rho$.}
\begin{ruledtabular}
\begin{tabular}{cc}
References     &   $K$ ($\mathrm{GeV}^3$) \\
\hline
Ref.~\cite{Bijker:2020tns} &  $\approx 0.024$ \\
Ref.~\cite{Ortiz-Pacheco:2023kjn} & $0.02377\pm0.00014$ \\
Ref.~\cite{Garcia-Tecocoatzi:2023btk} & $0.0254\pm0.0012$ or $0.0245\pm0.0023$\\
\end{tabular}
\end{ruledtabular}
\label{tab:kValue}
\end{table}

The wave function used in this work is the single Gauss function. Thus, the values of $\alpha_\rho$ and $\alpha_\lambda$ should contain the non-harmonic effects and may deviate from the values extracted from the hadron spectrum. Hence, $\alpha_\rho$ and $\alpha_\lambda$ should be treated as two independent parameters from the perspective of phenomenology. 
In this work, in order to avoid the uncertainty caused by the charmed hadrons or light hadrons, we only consider those decay processes that involve the bottom baryons to extract the values of these parameters. However, only the decay width of $\Xi_b^-\to \Lambda_b^0 \pi^-$ has been measured with which the parameters cannot be effectively restricted. As the result, the strong decays of the other bottom baryons, such as the $\Sigma_b^{(*)}\to \Lambda_b^0 \pi$~\cite{ParticleDataGroup:2024cfk} strong decay processes, are introduced to constrain the parameters.

The strong decay vertices of $\Sigma_b \to \Lambda_b^0\pi$ will appear in the hadronic weak decay pole terms that are not shown separately. Its transition amplitudes are calculated with the chiral quark model introduced in the Sec.~\ref{sec:framework}. 
Apparently, the available data for the strong decay width of $\Sigma_b^+\to \Lambda_b^0 \pi^+$ and $\Xi_b^-\to \Lambda_b^0 \pi^-$ are insufficient to determine the parameters $\alpha_\rho$, $\alpha_\lambda$, and $\delta$. So, we have to reconsider the relation given by Eq.~(\ref{eq:arholambda}) to reduce the parameters, and treat $\delta$ and $K$ as the free parameters to be determined in our framework.

The center values and the errors of these two parameters are extracted with Monte Carlo (MC) technique instead of fitting. The acceptance-rejection method~\cite{ParticleDataGroup:2024cfk} is used to generate samplings of $\delta$ and $K$. We assume that the distribution of the decay width of $\Sigma_b^+\to \Lambda_b^0 \pi^+$ and $\Xi_b^-\to \Lambda_b^0 \pi^-$ obeys the normal distribution. The sampling interval of $\delta$ is $(0.5,0.8)$ and the sampling interval of $K$ is $(0,0.18)~\mathrm{GeV}^3$.

\begin{figure}[htbp]
\begin{center}
\includegraphics[scale=0.6]{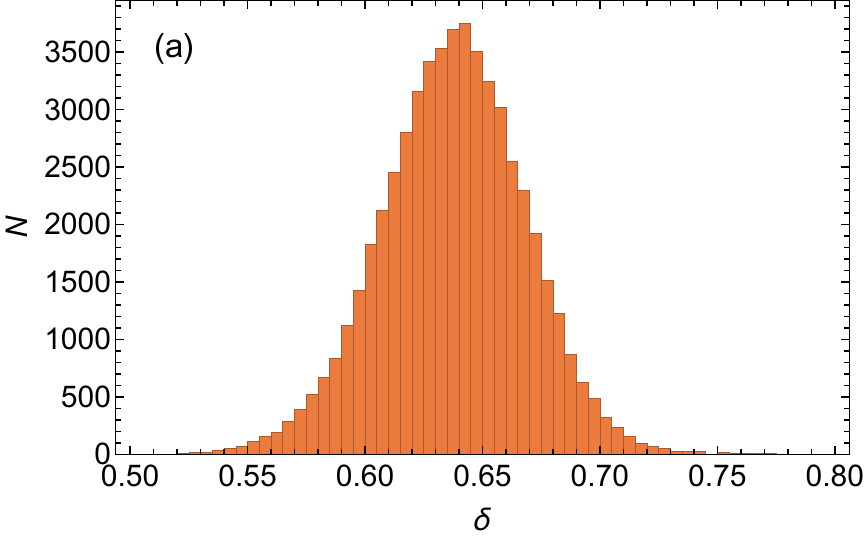}
\includegraphics[scale=0.6]{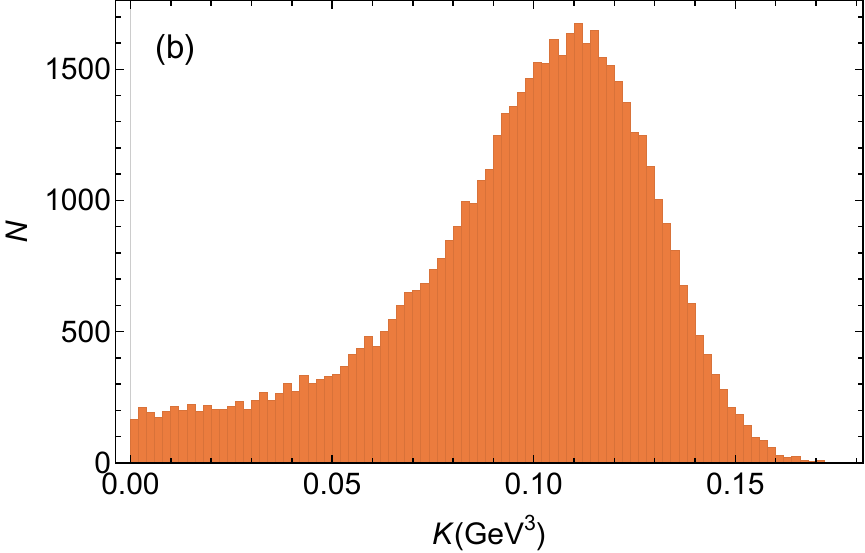}
\caption{The distribution of $\delta$ and $K$ constrained by the strong decay channel $\Sigma_b^+ \to \Lambda_b^0 \pi^+$ and hadronic weak decay channel $\Xi_b^- \to \Lambda_b^0 \pi^-$. The total number of the samples is about 54300.}
\label{fig:deltak}
\end{center}
\end{figure}

The distributions of the parameters $\delta$ and $K$ constrained by the strong decay of $\Sigma_b^+ \to \Lambda_b^0 \pi^+$ and hadronic weak decay of $\Xi_b^- \to \Lambda_b^0 \pi^-$ are shown by Fig.~\ref{fig:deltak}.  
The total number of the samples is about $5.4\times 10^4$ for each parameter. Fig.~\ref{fig:deltak}(a) clearly show that the distribution of $\delta$ obeys the normal distribution. The center value of $\delta$ is about $0.64$ that is lightly larger than the value used in Ref.~\cite{Zhong:2007gp}. The standard deviation is about $0.03$.
As shown in Fig.~\ref{fig:deltak}(b), the distribution of $K$ apparently deviates from the normal distribution. It may result from the lack of the input data. 
We take the maximum probability value  that is about $0.11~\mathrm{GeV}^3$ as the center value of $K$ and take the half-width of the distribution that is about $0.02~\mathrm{GeV}^3$ as the error of $K$. It should be noted that the center value of $K$ extracted in this work is significantly larger than the values listed in Tab.~\ref{tab:kValue}. In contrast, the corresponding $\alpha_\rho$ of $\Xi_b$ is about $0.57$ GeV that is about 1.4 times of $\alpha_\rho$ (0.40 GeV) used in Ref.~\cite{Bijker:2020tns}. This deviation can be ascribed to the approximation of the spatial wave function as mentioned before. As discussed earlier, the deviation of $\alpha_\rho$ is acceptable.

\subsection{The numerical results and discussion}

With the center values of $\delta$, $K$ and the quarks masses determined in the previous section, we calculate the amplitudes for each channel and the non-zero transition matrix elements are listed in Tab.~\ref{tab:amps}.
Depending on the distribution of $\delta$ and $K$, the center values and the errors of the calculated branching ratios of $\Xi_b \to \Lambda_b^0\pi$ and $\Omega_b \to \Xi_b \pi$ are listed in Tab.~\ref{tab:results}.

\begin{table*}[htbp]
\centering
\caption{The amplitudes for each processes (in unit of $10^{-9}~\mathrm{GeV}^{-1/2}$). The spin of the initial and final baryon is taken as $1/2$. `CS-I' and `CS-II' are used to label the first and second type of the CS process, respectively. }
\begin{ruledtabular}
\begin{tabular}{lcccc}
processes       & $\Xi^-_b \to \Lambda_b^0 \pi^-$ & $\Xi_b^0 \to \Lambda_b^0 \pi^0$ & $\Omega_b^- \to \Xi_b^0 \pi^-$  & $\Omega_b^- \to \Xi_b^- \pi^0$ \\
\hline 
 DPE                           &$-9.86$          &$0$            &$3.05$      &$0$\\
 CS I                          &$8.65$           &$6.12$         &$0$         &$0$   \\
 CS II                         &$0$              &$2.32$         &$0$         &$0.71$     \\
WS                             &$0$              &$0$            &$0$         &$0$    \\
SW($\Xi_b^0|^2P_\rho\rangle$)  &$-21.37-0.09i$   &$-14.91-0.06i$  &$0$         &$0$   \\
SW($\Xi_b^0|^4P_\rho\rangle$)  &$-27.58-0.07i$   &$-19.24-0.05i$  &$0$         &$0$   \\
 Total                         &$-50.16+0.16i$   &$-25.70-0.11i$  &$3.05$      &$0.71$    \\
\end{tabular}
\end{ruledtabular}
\label{tab:amps}
\end{table*}

\begin{table*}[htbp]
\centering
\caption{The branching ratios (in \%) of $\Xi_b \to \Lambda_b^0\pi$ and $\Omega_b \to \Xi_b \pi$ are compared with the experimental data and other models results. Our results are list in the last line. The error is related with the uncertainty of $\delta$ and $K$.}
\begin{ruledtabular}
\begin{tabular}{lllll}
References    & $\Xi_b^-\to \Lambda_b^0 \pi^-$      & $\Xi_b^0\to \Lambda_b^0 \pi^0$        
&$\Omega_b^-\to \Xi_b^0 \pi^-$ &$\Omega_b^-\to \Xi_b^- \pi^0$ \\
\hline
Experimental Data &$0.89\pm 0.46$~\cite{LHCb:2023tma} & $\cdots$ & $\cdots$ & $\cdots$ \\
Ref.~\cite{Cheng:2022jbr}  &$0.42\pm0.05$  & $0.26\pm0.03$   & $(6.5\pm0.7)\times10^{-3}$ & $(3.2\pm0.3)\times10^{-4}$\\
Ref.~\cite{Li:2014ada}  &$0.8\pm0.3$  & $\cdots$ & $\cdots$ & $\cdots$\\
Ref.~\cite{Gronau:2015jgh}  &$0.63\pm0.42$  & $0.32\pm0.21$  & $\cdots$ & $\cdots$\\
Ref.~\cite{Faller:2015oma}  &$0.19-0.76$  & $0.09-0.37$& $(1.1-4.3)\times 10^{-4}$ & $(0.6-2.2)\times 10^{-4}$\\
Ref.~\cite{Li:2014ada}  &$2-8$  & $1-4$ &$\cdots$ & $\cdots$\\
Previous Work~\cite{Niu:2021qcc} &$0.14 \pm0.073$ & $0.017\pm0.015$ & $\cdots$ &$\cdots$\\ 
This work &$0.92\pm0.40$  & $0.22\pm0.13$ & $(9.12\pm 0.07)\times10^{-3}$ & $(4.77\pm 0.035)\times10^{-4}$\\ 
\end{tabular}
\end{ruledtabular}
\label{tab:results}
\end{table*}

\subsubsection{The asymmetry parameter}

For these hadronic weak decay channels $\Xi_b\to \Lambda_b^0 \pi$ and $\Omega_b\to\Lambda_b^0\pi$, $b$ quark does not participate in interactions and behaves as a spectator. One would expect that the heavy quark spin symmetry (HQSS) should hold in these decay channels. As a consequence, one can describe the decay processes through the transitions between the light quarks, i.e.
$s_i^P\to s_{f_1}^{P_1}s_{f_2}^{P_2}$ with $s_i$ ($s_{f_{1,2}}$) and $P$ ($P_{1,2}$) denoting the spin and parity of the initial (final) light quarks (or quark-antiquark), respectively. With this convention, the hadronic weak decays of  $\Xi_b\to \Lambda_b^0 \pi$ with the light flavor of anti-triplet will be via $0^+\to 0^+0^-$. It shows that only the $S$-wave transition operator can conserve the angular momentum. Thus, the leading order amplitude should be given by the $S$-wave PV transition. Similarly, the decay of $\Omega_b\to\Lambda_b^0\pi$ hadronic weak decays with the light flavor of sextet will be via $1^+\to 0^+0^-$, and the leading order amplitude will be given by the $P$-wave PC transition.

The numerical results do show that the contributions of the PC operators to the process of $\Xi_b\to \Lambda_b^0 \pi$ and the contributions of the PV operators to the process of $\Omega_b\to\Lambda_b^0\pi$ vanish as we expected. This causes that the asymmetry parameter $\alpha$~\cite{ParticleDataGroup:2024cfk}, which is determined by the interferences between the PC and PV components of the transition amplitude
\begin{align}
\alpha=\frac{2 \operatorname{Re}\left[\mathcal M^*_\mathrm{P V} \mathcal M_\mathrm{P C}\right]}{\left|\mathcal M_\mathrm{P C}\right|^2 +\left|\mathcal M_\mathrm{P V}\right|^2 } \ ,
\end{align}
vanishes for both $\Xi_b\to \Lambda_b^0 \pi$ and $\Omega_b\to\Lambda_b^0\pi$.
This vanishing asymmetry parameter is crucial for testing the heavy quark symmetry in these two processes.

\subsubsection{$\Xi_b \to \Lambda_b^0\pi$}
Fig.~\ref{fig:Xibm} and Fig.~\ref{fig:Xibz} clearly reveal that the $\Xi_b^-\to \Lambda_b^0 \pi^-$ and $\Xi_b^0 \to \Lambda_b^0 \pi^0$ decay processes are not exactly the same. 
%This leads to the flavor symmetry is broken in the decay width level. 
In the following, we will compare the contributions of the DPE, CS processes and the pole terms to reveal the role played by flavor symmetry and the hadronic weak decay mechanisms for these two processes.

Due to the suppression of the neutral current interaction, we do not consider the DPE in $\Xi_b^0\to \Lambda_b^0 \pi^0$. The CS-I process can contribute to both $\Xi_b^-\to \Lambda_b^0 \pi^-$ and $\Xi_b^0 \to \Lambda_b^0 \pi^0$ processes. Comparing the calculated amplitudes from the CS-I process in Tab.~\ref{tab:amps}, one can find that the ratio of about 1.4 between these two channels satisfies the isospin symmetry. The weak transition operator of the CS-I process is $s\to u\bar u d$, and 
%its weak isospin is $|\Delta I,\Delta I_3\rangle=|\frac{1}{2},-\frac{1}{2}\rangle $. 
according to the weak isospin analysis, we have 
\begin{align}
\frac{\mathcal{A}_{\mathrm{CS-I}}(\Xi_b^-\to \Lambda_b^0 \pi^-)}{\mathcal{A}_{\mathrm{CS-I}}(\Xi_b^0 \to \Lambda_b^0 \pi^0)}
=\frac{\langle1,-1|\frac{1}{2},-\frac{1}{2};\frac{1}{2},-\frac{1}{2}\rangle}{\langle 1,0|\frac{1}{2},\frac{1}{2};\frac{1}{2},-\frac{1}{2}\rangle}=\sqrt 2.
\end{align}
This result can also be obtained from the transition diagrams directly. One can see that Fig.~\ref{fig:Xibm}(b) and Fig.~\ref{fig:Xibz}(a) share the same decay mechanism of the CS-I transition. In Fig.~\ref{fig:Xibz}(a) the projection of the $u\bar u$ component to the $\pi^0$ wave function will contribute a factor of $1/\sqrt 2$.

The CS-II process only contributes to the decay of $\Xi_b^0 \to \Lambda_b^0 \pi^0$. The amplitude of the CS-II process is smaller compared with the CS-I process,  as shown in Tab.~\ref{tab:amps}. It is because of the difference arising from the convolution of the spatial wave functions.

As shown in Tab.~\ref{tab:cbmass} and \ref{tab:amps}, we see that most of the states have no contributions to the pole terms. Only the two states, $\Xi_b^0(|^2P_\rho\rangle)$ and $\Xi_b^0(|^4P_\rho\rangle)$, contribute to the SW pole terms. It indicates that the transitions between the bottom baryons are subjected to strict restrictions because of the HQSS or the conservation of isospin and the relevant selection rules as discussed in Ref.~\cite{Niu:2021qcc}.

The ratio between the amplitudes of SW pole terms  for $\Xi_b^-\to \Lambda_b^0 \pi^-$ and $\Xi_b^0 \to \Lambda_b^0 \pi^0$ is also about $1.4$, which is given by the isospin factors for the $\pi^-$ and $\pi^0$  production at the strong interaction vertices. 
%conservation of the strong interaction that can be easily verified. 
It can be recognized from  Fig.~\ref{fig:Xibm}(c) and  Fig.~\ref{fig:Xibz}(e) which share the same transition mechanism except that the  $u\bar u$ component will project to the $\pi^0$ flavor wave function and contribute a factor of $1/\sqrt{2}$.

Notice that the amplitudes of the SW pole terms is much larger than the DPE and CS processes in both $\Xi_b^-\to \Lambda_b^0 \pi^-$ and $\Xi_b^0 \to \Lambda_b^0 \pi^0$. It suggests  that the pole terms should play a crucial role in understanding the decay property of these two processes. Even with the interference effects among all the amplitudes taken into account, the ratio of about 2 between their total amplitudes does not deviate too much from $\sqrt{2}$ (or numerically 1.4) given by the isospin factor. It results in the branching ratio fraction between $\Xi_b^-\to \Lambda_b^0 \pi^-$ and $\Xi_b^0 \to \Lambda_b^0 \pi^0$ a factor of about 4, which is in agreement with the experimental measurement as shown in Tab.~\ref{tab:results}. 
In contrast, if we neglect the contributions of the factorizable process, i.e. the DPE contribution to $\Xi_b^-\to \Lambda_b^0 \pi^-$, we will have 
\begin{align} \label{eq:relation}
\mathrm{Br} (\Xi_b^-\to \Lambda_b^0 \pi^-)\approx 2 \mathrm{Br} (\Xi_b^0\to \Lambda_b^0 \pi^0) \ ,
\end{align}
which agrees with the prediction of Ref.~\cite{Cheng:2022jbr}. In  Ref.~\cite{Cheng:2022jbr}, it is argued that the contribution from the factorizable term should be small and the relation of Eq.~(\ref{eq:relation}) will hold approximately.

 At the end, we compare our results with others in Tab.~\ref{tab:results}, from which one can see that most of other works underestimate the experimental data. Our previous result for the branching ratio is $(0.14\pm 0.073)\%$ for the $\Xi_b^-\to\Lambda_b^0 \pi$~\cite{Niu:2021qcc}, which is also much smaller than the experimental data. There, the value of $K=0.035~\mathrm{GeV}^3$ is adopted  based on the values extracted from the heavy baryon spectra and is much smaller than the value used in this work. As pointed out in Ref.~\cite{Niu:2021qcc} the amplitudes is sensitive to the value of $K$ and can be different for hadrons with different flavors. In this work, instead, the value of $K$ is constrained by the experimental data for   $\Xi_b^- \to \Lambda_b^0 \pi ^-$, with which one can predict the branching ratios of othepredictions for other channels can be made. Because of this, the branching ratio of  $\Xi_b^0\to \Lambda_b^0 \pi^0$ obtained in this work is different from the previous work~\cite{Niu:2021qcc}, but should be more self-consistent here.

\subsubsection{$\Omega_b \to \Xi_b \pi$}

For the decay process of $\Omega_b^-\to \Xi_b^0 \pi^-$, the tree-level transition is via the DPE process as shown by Fig.~\ref{fig:Omegapim}, and for $\Omega_b^-\to \Xi_b^- \pi^0$ the tree-level transition is via the CS-II process as shown by Fig.~\ref{fig:Omegapiz}. It should be noted that no pole terms can contribute to these two processes that leads to two obvious features for the $\Omega_b^-\to \Xi_b^0 \pi^-$.

Firstly, this makes it hard to relate these two transitions to each other directly through isospin symmetry. However, due to the effects of color suppression, the branching ratio of $\Omega_b^-\to \Xi_b^- \pi^0$ is supposed to  be much smaller than that of $\Omega_b^-\to \Xi_b^0 \pi^-$ by about one order of magnitude. 
Secondly, the branching ratios of $\Omega_b\to \Xi_b \pi$ is much smaller than those of $\Xi_b\to \Lambda_b^0\pi$ as shown in Tab.~\ref{tab:results}, which is because of the absence of the pole terms.
More specifically,  this is because $\Omega_b^-\to \Xi_b^0 \pi^-$ proceeds only through the DPE process that is proportional to the momentum of the final state which is about 205 MeV. It leads to the suppression of the DPE amplitude which is consistent with the conclusion of Ref.~\cite{Cheng:2022jbr}.
In addition, as clearly shown by Tab.~\ref{tab:amps}, the  amplitude of the DPE process of $\Omega_b^-\to \Xi_b^0 \pi^-$ and the CS-II amplitude of $\Omega_b^-\to \Xi_b^- \pi^0$ are much smaller than the pole terms of $\Xi_b\to \Lambda_b^0\pi$. Thus, the decay width of $\Omega_b^-\to \Xi_b^- \pi^-$ is about two orders of magnitude smaller than $\Xi_b^{-} \to \Lambda_b^0 \pi^-$ although the phase space of $\Omega_b \to \Xi_b \pi$ is relatively larger.

Before the conclusion, we would like to discuss the final-states interaction in bottom baryon decays. 
The role played by the final-states interaction has aroused much concern about the branching ratio anomaly and CP violation effects in hadronic weak decay of heavy baryons~\cite{Han:2021azw, Jia:2024pyb,  Duan:2024zjv, Hu:2024uia, Cheng:2025oyr}.
It is generally accepted that the final-state interaction plays an essential role in the study of exotic states near threshold~\cite{Guo:2017jvc,Guo:2019twa, Dong:2020hxe}. As we have learned in many cases, the final-state interaction can become important when the interaction is an $S$-wave and near threshold. At the leading order, only the $P$ wave contributes to the $\Omega_b\to \Xi_b\pi$ process~\cite{Liu:2011xc,Song:2025nxi}. This leads to the final-state interaction less important since  the HFC hadronic weak decay is a near-threshold processes. So the effects of the final-state interaction are neglected at this moment.

\section{Summary and Outlook}

The HFC hadronic weak decays of $\Xi_b \to \Lambda_b^0\pi$ and $\Omega_b \to \Xi_b\pi$ are investigated in the framework of the NRCQM. 
In our calculation, the values of the quark masses are constant with the commonly adopted values.
The values of the spring constant $K$ and the dimensionless parameter $\delta$ are extracted from the measured decay widths of $\Xi_b^- \to \Lambda_b^0\pi^-$ and $\Sigma_b^+\to \Lambda_b^0\pi^+$. 
The value of $K$ extracted in this work is different from that given by the hadron spectrum study. This may result from the harmonic oscillator approximation, which is an oversimplified approach in our framework. Since the systematic model uncertainties can be absorbed into some of the model parameters, the advantage of the NRCQM framework is to treat the DPE, CS and pole terms in a self-consistent way and their relative strengths can be reliably evaluated. 
%can be calculated simultaneously in framework of NRCQM.

With the contributions of the DPE, CS and pole terms systematically considered, the numerical results show that these mechanisms manifest themselves differently and are crucial for understanding the branching ratio fraction between $\Xi_b^- \to \Lambda_b^0\pi^-$ and $\Sigma_b^+\to \Lambda_b^0\pi^+$. 
It shows that the contributions of the pole terms can be significantly larger than the contributions of the DPE and CS processes. This is analogous with the case of the $\Xi_c\to\Lambda_c \pi$. Our analysis also shows that the decay width of $\Omega_b\to \Xi_b \pi$ is significantly smaller than that of $\Xi_b \to \Lambda_b^0\pi$ due to the absence of the pole terms. The prediction can be examined by the future experiment at LHCb and Belle-II.

%%%%%%%%%%%%%%%%%%%%%%%%%%%%%%%%%%%%%%%%%%%%%%%%%%%%%%%%%%%%%%%%%%%%%%%%%%%%%%%%
\begin{acknowledgements}
This work is supported, in part, by the National Natural Science Foundation of China with Grant Nos.~12505110, 12235018, 12547105, and 12375073. 

\end{acknowledgements}

%%%%%%%%%%%%%%%%%%%%%%%%%%%%%%%%%%%%%%%%%%%%%%%%%%%%%%%%%%%%%%%%%%%%%%%%%%%%%%%%
\begin{appendix}

\end{appendix}

%\bibliography{ref.bib}

\end{document}